\documentclass[floatfix,twocolumn,english,showpacs,preprintnumbers,superscriptaddress,prl,checklength]{revtex4}
\usepackage{ae,aecompl}
\usepackage[T1]{fontenc}
\usepackage[latin1]{inputenc}
\usepackage{verbatim}
\usepackage{textcomp}
\usepackage{amsmath,amsfonts}
\usepackage{graphicx}
\usepackage{amssymb}
\usepackage{multirow}
\makeatletter

\usepackage{subfigure}
\usepackage{dcolumn}\usepackage{bm}\@ifundefined{definecolor}
 {\usepackage{color}}{}
\usepackage{latexsym}

\makeatother

\usepackage{babel}
\definecolor{Red}{rgb}{1,0.0,0.0}
\begin{document}

\title{Experimental protection of quantum gates against decoherence }
\title{Experimental protection of quantum gates against decoherence and control errors }

\author{Alexandre M. Souza}

\author{Gonzalo A. \'{A}lvarez}

\author{Dieter Suter}

\affiliation{Fakult\"at Physik, Technische Universit\"at Dortmund, D-44221,
Dortmund, Germany}

\date{\today}
\begin{abstract}
One of the biggest challenges for implementing quantum devices is the requirement
to perform accurate quantum gates.
The destructive effects of interactions with the
environment present some of the most difficult obstacles that must be overcome for  precise quantum control.
In this work we implement a proof of principle  experiment of quantum gates protected against a fluctuating environment
 using dynamical decoupling  techniques.
We show that  decoherence can be reduced during the application of quantum gates.
High fidelity quantum
gates can be achieved even if the gate time exceeds the decoherence time by one order of magnitude.
\end{abstract}

\maketitle

\emph{Introduction.---}
 Quantum information processing (QIP) \cite{nielsen} can lead to a
dramatic computational speed-up over classical
computers for certain problems.
However, any physical QIP device is subject to errors
arising from  unavoidable interactions with the
environment or from control imperfections.
Therefore, scalable QIP  needs
methods for avoiding or correcting those errors.
The theory of quantum error
correction (QEC) states that it is possible to stabilize a quantum computation provided that the
error per gate is below some threshold \cite{zurek98} and high fidelity initial
states are prepared \cite{kitaev2005,laflamme2010}.
However, QEC needs  many
auxiliary qubits, generating a significant overhead in additional resources.  Therefore, it is highly desirable to
 develop methods for
reducing the perturbation both between and inside the quantum gates
without requiring additional  qubits.
In addition, reaching the error threshold required for the implementation of QEC requires
that the fidelity of the individual gate operations must be very high \cite{ladd}.

A simple way to avoid decoherence and thus reduce the error per gate consists in
 choosing qubit systems whose decoherence
times are long compared to the duration of a gate operation.
However, this is not always possible; in particular in systems that combine different types of qubits,
such as electronic and nuclear spins, the decoherence of the electron spin can be faster than the
possible gate durations of the nuclear spins. Furthermore, the reduction of the decoherence time of a
quantum register with the number of qubits \cite{Krojanski1,Krojanski2} imposes an additional
difficulty for implementing  quantum computation in large systems.

Dynamical decoupling (DD) \cite{viola99,yang,suter2012} is a promising method developed
to reduce decoherence by attenuating the
system-environment interaction with a sequence
of inversion pulses periodically applied to the qubits.  Recent experiments
have successfully implemented
DD methods and demonstrated the resulting increase
of the storage times in different
systems \cite{bollinger2009,liu2009,suter2010,hanson2010,cory2011,suter2011,suter2011b}.
In all these
implementations, the goal was to preserve a given input state, i.e. to protect a quantum memory
against environmental perturbations.
The necessity of protecting qubits against environmental noise occurs also in the context of
quantum information processing.
In a structured environment that induces well-characterized relaxation pathways, it is possible
to design protected gate operations by optimal control techniques \cite{glaser2011}.
If the relaxation mechanism is not known or it affects all modes of the system,
it may still be possible to use DD techniques, provided the effect of the DD sequence
is compatible with the gate operations used for information processing.
In the simplest case, gate operations can be made insensitive to static environmental perturbations
by refocusing them  \cite{Burum1980, yacoby2012}, in a manner quite similar to a Hahn echo.
In the more general case of a fluctuating environment, the Hahn echo has to be replaced by
multiple-pulse DD sequences.
Initial experiments in this direction have been made recently
by applying gate operations \emph{between} cycles of DD sequences acting on the electron spin qubit
of an NV center \cite{dobrovitski2012} or  an effective qubit in a semiconductor quantum dot \cite{gossard2010}.
In these schemes, the DD protection is not active during the actual gate operation.
Possible scheme for maintaining DD protection \emph{during} the gate operation were also suggested
\cite{viola2009,viola2010,gyure2010,viola2009b,preskill2011,lukin2009}.
These schemes were developed under the assumption of perfect controls,
i.e. the controls operations used e.g. for DD should not introduce any additional errors.
Here, we propose a general scheme for protection against a fluctuating environment
that does not rely on this assumption but is robust against experimental errors and can therefore
be implemented in an experimental scheme with realistic control operations.

We show that the
decoherence can be reduced during the application of quantum gates for a single qubit in a solid state system.
High fidelity quantum
gates are achieved even if the gate time exceeds the decoherence time by one order of magnitude.
Since the protection scheme introduces many additional control operations,
we design the protected gate operations in such a way that the effect of control imperfections
on the fidelity of the system is minimized.

\emph{Protection scheme.---}
Quantum logical gates are achieved by using time-varying
control Hamiltonians  $H_c(t)$ for the relevant qubit system.
Their propagators can be represented by unitary operators
\begin{eqnarray}
U = \mathcal{T} e^{-i \int_0^T H_c(t) dt} = e^{-i H_G T}
\label{u}
\end{eqnarray}
where $\mathcal{T}$ is the Dyson time ordering operator and $T$ is the gate time.
The propagator can also be expressed in terms of the  time-independent average Hamiltonian $H_G$,
which would generate the same operation if it were active for the time $T$.
For single qubit gates, this Hamiltonian can be expressed as:
\begin{eqnarray}
H_G = \omega  \, \vec{n}  \cdot \mathbf{S}
\label{hg}
\end{eqnarray}
where $\mathbf{S} = (S_x,S_y,S_z)$ is the spin vector operator
of the system qubit, $\vec{n} $ is a 3D vector and the strength $\omega$ is a real parameter.

In any physical implementation, the system also
interacts with the environment, which introduces
decoherence reducing the gate fidelity.
If the environmental effects become
too strong, the quantum computation cannot be stabilized by QEC codes \cite{zurek98}.
One approach to avoid this is to use DD for reducing the effects of the environment.
Consider the Hamiltonian describing a one qubit system and its environment
\begin{eqnarray}
H = H_S + H_{SE} + H_E,
\label{HH}
\end{eqnarray}
where $H_S = \omega_S S_z $ is the system Hamiltonian and $\omega_S$ is the Larmor
frequency of the system. $H_E$ is the environmental
Hamiltonian and $H_{SE}$ the system-environment coupling.
Here, we describe the environment as a spin bath and the coupling as a pure dephasing interaction
\begin{eqnarray}
H_{SE} &=&  \sum_k b_k S_z I_z^k ,
\end{eqnarray}
where $I^k_z$ is
the spin operator
of the $k^{th}$ environment spin,
$b_k$  is the coupling constant between the system
and the $k^{th}$ spin of the environment.

The identity-operation ($H_G = 0$) can be implemented by just applying
DD sequences, such as the XY-4 sequence.
This sequence was
initially introduced in the context of
nuclear magnetic resonance (NMR)  \cite{Maudsley1986,conradi}.
A similar sequence is the PDD sequence \cite{viola99,suter2011,suter2012b}.
Neglecting pulse errors, we can write
the zeroth and first order terms of the average
Hamiltonian of the XY-4 sequence as $\overline{H_0} = H_E$ and  $\overline{H_1} = 0$. The
system-environment Hamiltonian, which causes decoherence, as well as
the internal Zeeman Hamiltonian
of the  system qubit are removed up to the first order approximation.
The only remaining term  is the environmental Hamiltonian $H_E$, which has no
effect on the system qubit.

\begin{figure}[htbp]
\vspace*{13pt}
\begin{center}
{\includegraphics[width=9.5cm]{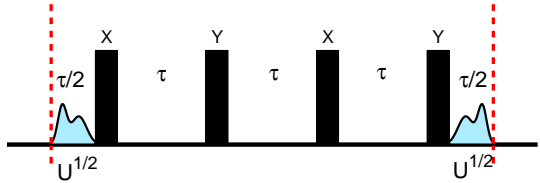}}
\end{center}
\vspace*{13pt}
\caption{\label{f1} Pulse sequence for  protecting quantum gates using the XY-4 sequence. }
\end{figure}

A simple way to introduce gates with real pulses during decoupling consists in dividing the
gate into two equal parts  $\sqrt{U} = e^{-i H_G T/2-i H T/2}$.
Here, $e^{-i H_G T}$ is the target control operation, while the drift term $H T/2$ is an unwanted
contribution that we cancel in the protected control operation.
We insert the two half-gates $\sqrt{U}$ into the initial and final
free precession periods of the DD sequence, as shown in Fig. \ref{f1}.
Adjusting $T = \tau$, the delay between the DD pulses, the propagator for the full cycle
(duration $\tau_c = 4 \tau$, neglecting the pulse duration) becomes, to first order in the cycle time,
\begin{eqnarray}
U & \approx &  e^{-4 i H_E  \tau} e^{-i  H_G  T} .
\label{propag}
\end{eqnarray}

The free evolution terms $\propto H_S+H_{SE}$ are cancelled to first order by DD.
This scheme resembles the theoretical approach of Refs. \cite{viola2009,viola2009b},
except that our scheme has higher symmetry, which helps to eliminate some control errors \cite{suter2012b}
.
The remaining terms are $H_E$, which acts only on the environment,
and the gate operator $H_G$.
Since the effect of the SE-coupling is absent here, decoherence has been eliminated in this first order approximation.

\emph{Robust implementation.---}
In real implementations, experimental imperfections must also be taken into account.
In most cases, the dominant imperfection is a deviation between the
actual and the ideal amplitude of the control field. The result of this amplitude error
is that the rotation angle deviates from the target angle typically by a few
percent. The imperfect control can affect  both
the implementation of $H_G$ as well the $\pi$ pulses of the dynamical
decoupling sequence and its effect can be particularly devastating when the number
of gate operations is large and the errors accumulate.

The systematic control errors of the DD pulses can be reduced by
choosing robust DD sequences \cite{suter2011,suter2012}.
A well established method for eliminating control field errors is the use of composite pulses \cite{levitt}.
Composite pulses are sequences of consecutive pulses designed such that the resulting total operation
remains close to the ideal target operation even in the presence of some experimental imperfections.
A good choice for correcting amplitude errors in a general single qubit rotation $\mathbf{R}_{\phi}(\theta)$
is the  $BB1$ composite pulse \cite{chuang2004}:
\begin{eqnarray}
\mathbf{R}_{\phi}(\theta) = R_{\phi}(\theta/2)R_{\phi + \psi}(\pi)R_{\phi + 3\psi}(2\pi)R_{\phi + \psi}(\pi)R_{\phi}(\theta/2),\nonumber
\end{eqnarray}
where $\phi$ describes the rotation axis, $\theta$ the rotation angle, and $\cos \psi = - \theta /4 \pi$.

\begin{figure*}[htbp]
\vspace*{13pt}
\begin{center}
{\includegraphics[width=12.0cm]{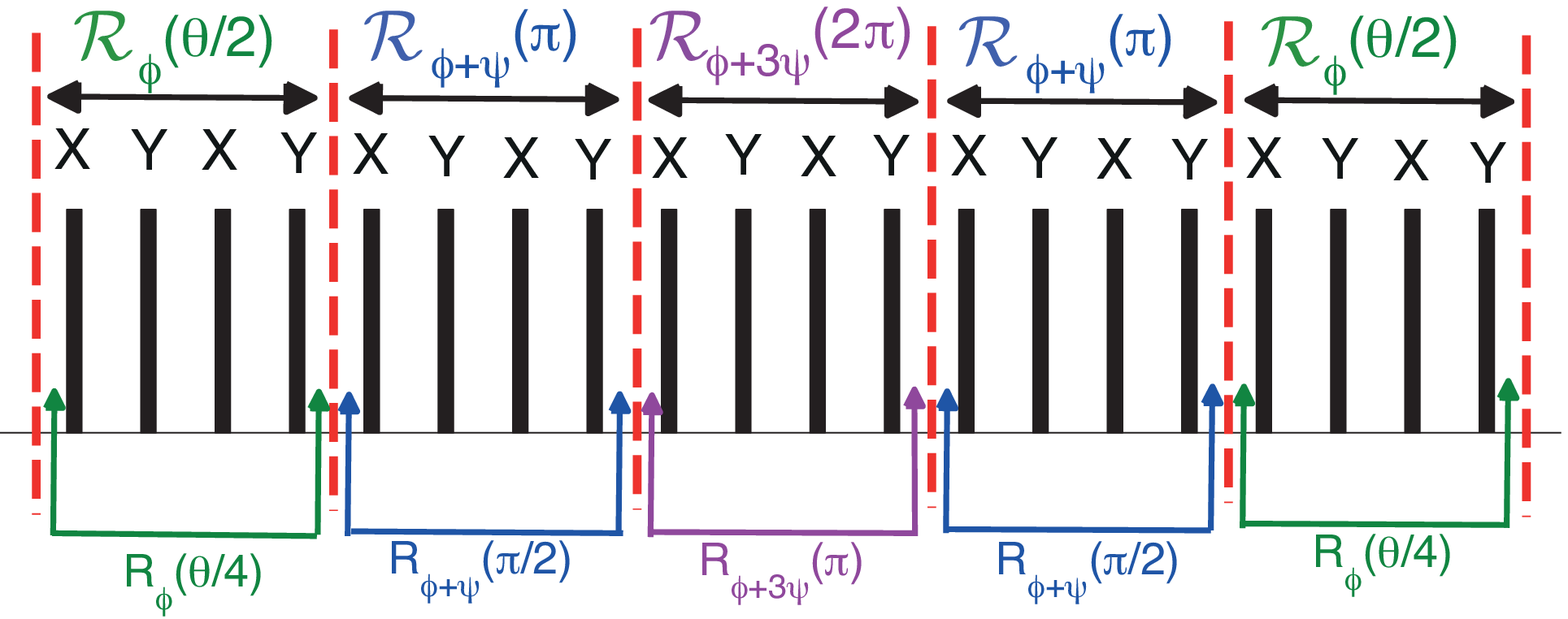}}
\end{center}
\vspace*{13pt}
\caption{\label{bb1gate} Pulse sequence for a decoherence-protected BB1 pulse
using the XY-4 sequence. }
\end{figure*}

Figure \ref{bb1gate} shows how a general
rotation can be made robust against amplitude errors and
protected against decoherence from a fluctuating environment.
For this purpose, we replace each rotation  $R_{\phi}(\theta)$ in the BB1 pulse
by the protected  rotation $\mathcal{R}_{\phi}(\theta)$ according to the scheme of Fig. \ref{f1}.
This scheme can obviously be extended to other DD sequence with symmetrical timing simply by
replacing the XY-4 cycle with a different cycle.

\begin{figure}[hbtp]
\vspace{13pt}
\begin{center}
{\includegraphics[width=9.5cm]{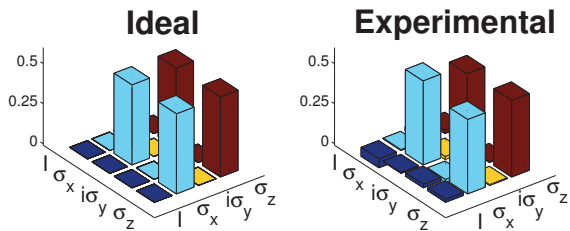}}
\end{center}
\vspace{13pt}
\caption{\label{tomo} { Process tomography of  the Hadamard gate.
The panels show the process matrix $\chi$ for an ideal gate and the
experimental process matrix of the protected gate.}
}
\end{figure}

\begin{figure}[htbp]
\vspace*{13pt}
\begin{center}
{\includegraphics[width=9.0cm]{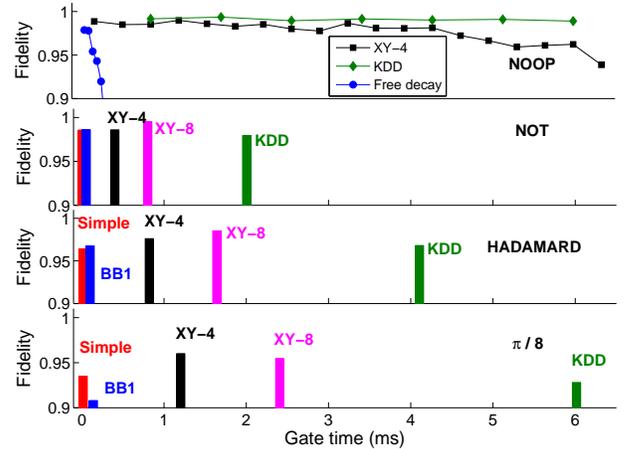}}
\end{center}
\vspace*{13pt}
\caption{\label{fidel} Gate fidelity as a function of gate time for different
gate operations protected by different DD sequences.
``Simple" indicates gates that were implemented by
unprotected rotations.  The delay between the pulses for the NOOP was $\approx 3 \mu$s. }
\end{figure}

\emph{Experimental performance.---}
For the experimental tests we used natural abundance
$^{13}$C nuclear spins in the CH$_2$ groups of a polycrystalline
adamantane sample as the system qubit.
In this system, the carbon
spins are coupled to $^1$H nuclear spins by heteronuclear
magnetic dipole interactions.
The protons are coupled to each other by  homonuclear dipolar interactions.
Under our conditions,
the couplings between the carbon nuclei can
be neglected. The experiments were performed on a homebuilt 300
MHz solid-state NMR spectrometer.

In the context of quantum information processing, it is important that the
performance of gate operations
be independent of the initial conditions. For quantifying the performance of a general
quantum operation, the fidelity $F$ can be used \cite{Wang2008}:
\begin{equation}
F = \frac{|Tr(A B^{\dagger})|}{\sqrt{Tr(A A^{\dag}) Tr(B B^{\dag})}}.
\label{e:fid}
\end{equation}
Here, $A$ is the target propagator for the process and $B$ the actual propagator.

The actual operations are not always unitary.
We therefore write the process as
\begin{equation}
\rho_f = \sum_{nm} \, \chi_{mn}  \,E_m  \,\rho_i  \, E_n^{\dag} ,
\end{equation}
where $\rho_i$ and  $\rho_f$ are the density matrices at the beginning and end of the process.
The operators $E_m$ must form a basis. For the present case, we choose them as
$E_m = ( I,\sigma_x,i\sigma_y,\sigma_z )$, with the Pauli matrices $\sigma_{\alpha}$.
The ideal and actual processes can therefore be quantified by
the matrix elements $\chi_{mn}$ and experimentally determined by quantum process tomography {\bf  \cite{Chuang1997}}.

In the experimental implementation of this concept, we tested the Hadamard (H), NOT
and $\pi / 8$ gates.
For protection, we used the DD sequences XY-4 \cite{Maudsley1986}, XY-8 \cite{conradi} and KDD \cite{suter2011}.
The gates were decomposed into sequences of rotations around axes in the $xy$-plane (see table  \ref{tab1})
and each rotation was implemented as shown in Fig. \ref{bb1gate}.
Fig. \ref{tomo}  shows the results of the quantum process tomography
for the case of the Hadamard gate protected by the XY-8 sequence.
Without dynamical decoupling, the system coherence decays on a timescale
$T_2^*\approx 370 \, \mathrm{\mu s}$. The decay time only due to  the interaction with a fluctuating environment
(measured by the Hahn echo) is $T_2\approx 750 \, \mathrm{\mu s}$.
As shown in Table \ref{tab1}, the gate fidelities  for XY-8
are $>0.95$, although the durations of the gate operations are $\gtrsim T_2$. All the fidelities were calculated from the process matrices obtained directly from the raw data, without optimization methods as used in previous experiments (see for example \cite{cory}).
The obtained fidelity values close to one indicate that  the accumulation of incoherent errors is well compensated even for a very large number of pulses.

\begin{table}
\begin{centering}
{\footnotesize }\begin{tabular}{cccc}

 &  \textbf{\footnotesize Rotations}{\footnotesize{} } & \textbf{\footnotesize Gate time (ms)}&\textbf{\footnotesize Fidelity}\tabularnewline
\hline
 &  & \tabularnewline
{\footnotesize \bf{H} } & {\footnotesize \bf{$R_x(\pi) R_y(\frac{\pi}{2})$} } & {\footnotesize \bf{$1.6$} }& {\footnotesize \bf{$0.985$} }\tabularnewline
 &  & \tabularnewline
\hline
&  & \tabularnewline
{\footnotesize \bf{NOT} } & {\footnotesize \bf{$R_x(\pi)$} } & {\footnotesize \bf{$0.6$} }& {\footnotesize \bf{$0.995$} }\tabularnewline
 &  & \tabularnewline
\hline

&  & \tabularnewline
{\footnotesize \bf{$\pi/8$} } & {\footnotesize \bf{$R_x(\frac{\pi}{2})R_y(\frac{\pi}{4})R_x(-\frac{\pi}{2})$} } & {\footnotesize \bf{$2.2$} } & {\footnotesize \bf{$0.955$} } \tabularnewline
 &  & \tabularnewline
\hline

\hline
\end{tabular}
\par\end{centering}{\footnotesize \par}

\caption{{Implemented quantum gates. The gate times and  fidelities refer to the
experiment in which the XY-8 cycle is used.    \label{tab1} }}
\end{table}

In Fig. \ref{fidel}, we show the achieved gate fidelities for  four  gate operations:
identity = NOOP, NOT, Hadamard and $\pi/8$ phase gate.
For each type of gate, we plot the achieved fidelity against the total operation time for the gate,
using the direct implementation, labeled `simple' in the figure, using only BB1 pulses and protected gates with
different DD sequences.
The three DD sequences differ in cycle time and robustness, with XY-4 having the shortest cycle
and KDD being the most robust sequence  \cite{suter2011}.
In the top panel, we plot the fidelity of the NOOP gate for different gate durations,
while the lower three panels only show the fidelity for the shortest cycle with each
DD sequence.
In all cases, the measured gate fidelities were  $>0.928$.
This result is very gratifying, since it shows that high gate fidelities can be obtained even if
the gate duration exceeds the decoherence time  ($T^\ast_2$)  by an order of magnitude.
Additionally, they exceed the $T_2$ time, given by the Hahn echo decay, being a clear indication that the fluctuating environment  has been decoupled during the gate execution.

\emph{Discussion and Conclusion.---}
In summary, we have presented a proof of principle demonstration of
decoherence suppression during quantum logical gate operations
by dynamical decoupling.
For this purpose, we inserted robust gate operations  into different DD sequences
in such a way that they do not interfere destructively with the DD.
Using quantum process tomography, we have shown that
high fidelity single qubit quantum gates can be achieved even if the gate time exceeds
the decoherence time by one order of magnitude.
We carefully designed the protection scheme to be robust against deviations of the
control fields.
As a result, even for protected operations consisting of up to 330 individual control pulses,
the resulting fidelity is not significantly reduced compared to a gate implemented with a single pulse,
and often it is higher.
For some systems, slower gates promise higher fidelities than short gates \cite{slow1,slow2}.
In these cases, the approach that we have demonstrated here, appears particularly appealing
for further increasing the robustness and precision of the gate operations.

This result indicates that quantum
computation can be made reliable even for  systems in which the gate time is comparable to
or even greater than the decoherence time of the individual qubits.
In the present study, we have tested the scheme with three different robust DD sequences.
Our results are an additional demonstration
that dynamical decoupling  can be a useful tool that complements
quantum error correction.
We expect that the scheme is equally applicable to other types of qubit systems
as well as to other types of gate operations.
In particular, it will be interesting to apply this concept also to  multi-qubit systems.
In this case we need to refocus the undesired system-environment interactions
without eliminating the desired qubit-qubit interactions.
This leads to more complex pulse sequences but can be achieved in principle.
Efficient methods for selectively turning ``on" and  ``off"  specific
 Hamiltonian terms have been proposed in \cite{leung,mahler2001}.
While the evolution-time overhead grows linearly in these schemes,
a particular scheme \cite{yamaguchi}  designed for a networks of dipolar-coupled spins
leads to an evolution-time that is independent of the number of qubits.
A proposal to combine this scheme with dynamical decoupling was made in \cite{kern}.

\begin{acknowledgments}
We acknowledge useful discussions with Daniel Lidar.
This work is supported by the DFG through Su 192/24-1.
\end{acknowledgments}

\bibliography{gate}

\end{document}